\documentstyle[12pt]{article}
\def\be{\begin{equation}}
\def\ee{\end{equation}}

\newtheorem{theorem}{Theorem}

\newtheorem{definition}[theorem]{Definition}
\newtheorem{remark}[theorem]{Remark}

\begin{document}

\title{The Contextual Quantization and \\ the Principle of Complementarity of Probabilities}

\author{Andrei Khrennikov\footnote{Supported in part by the EU Human Potential Programme,
contact HPRN--CT--2002--00279 (Network on Quantum Probability and
Applications) and Profile Math. Modelling in Physics and Cogn.Sci.
of V\"axj\"o Univ.}
\\International Center for Mathematical
Modelling \\ in Physics and Cognitive Sciences,\\
University of V\"axj\"o, S-35195, Sweden, \\ e--mail: {\tt
Andrei.Khrennikov@msi.vxu.se}
\vspace{3mm} \\
Sergei Kozyrev\footnote{Supported in part by the EU Human
Potential Programme, contact HPRN--CT--2002--00279 (Network on
Quantum Probability and Applications) and Profile Math. Modelling
in Physics and Cogn.Sci. of V\"axj\"o Univ., by INTAS YSF
2002--160 F2, CRDF (grant UM1--2421--KV--02), The Russian
Foundation for Basic Research (project 02--01--01084) and by the
grant of the President of Russian Federation
for the support of scientific schools NSh 1542.2003.1.}\\
Institute of Chemical Physics,\\  Russian Academy of Science,
Moscow, Russia, \\ e--mail: {\tt kozyrev@mi.ras.ru}}

\maketitle

\begin{abstract}
The contextual probabilistic quantization procedure is formulated.
This approach to quantization has much broader field of
applications, compared with the canonical quantization. The
contextual probabilistic quantization procedure is based on the
notions of probability context and the Principle of
Complementarity of Probabilities. The general definition of
probability context is given. The Principle of Complementarity of
Probabilities, which combines the ideas of the Bohr
complementarity principle and the technique of noncommutative
probability, is introduced. The Principle of Complementarity of
Probabilities is the criterion of possibility of the contextual
quantization.
\end{abstract}

\section{Introduction}

The canonical approach to quantization (originated by Heisenberg's
''matrix mechanics'' \cite{Heisenberg}) is based, in essence, on
purely algebraic ideas. But quantum mechanics is a probabilistic
theory, see for discussion \cite{Holevo}, \cite{Ballentine},
\cite{KHR6}. Therefore it would be natural to create a
quantization procedure by starting directly with the probabilistic
structure of an experiment. Of course the authors are well aware
that a probabilistic structure can be added to the Heisenberg
algebraic quantization through the representation of
noncommutative Heisenberg variables by operators in a Hilbert
space and by using Born's probabilistic interpretation of
normalized vectors. P.Dirac developed ideas of W.Heisenberg and
created the modern variant of canonical quantization: mapping
dynamical variables to quantum observables and Poisson bracket to
commutator. Intrinsically Dirac's canonical quantization is also
an algebraic quantization. Probabilistic structure is added via
the Hilbert space representation of quantum observables.

In this paper we formulate the purely probabilistic approach to
quantization, based on the ideas of the Bohr complementarity
principle and the methods of noncommutative probability theory,
especially on the approach of contextual probability. For this
reason we formulate here the new general definition of
probabilistic context as a map of (in general, noncommutative)
probability spaces. We find that this new approach to quantization
is applicable for considerably broader class of models, compared
with the canonical quantization.

Let us remind the algebraic version of canonical approach to
quantization. Assume that we have a family of noncommutative
involutive algebra ${\cal B}(h)$ over complex numbers with the
commutation relations, which depend (say polynomially) on the
number parameter $h$, and for $h=0$ the algebra ${\cal B}(0)$ is
commutative.

Let us have the commutative Poissonian $*$--algebra ${\cal A}$
over the complex numbers, which is isomorphic to ${\cal B}(h)$ as
a linear space, and is isomorphic to ${\cal B}(0)$ as a
commutative $*$--algebra.

Therefore the mentioned isomorphism of linear spaces,
which relates the algebras ${\cal A}$ and ${\cal B}(h)$,
defines the $h$--dependent family of algebras ${\cal B}(h)$
with the elements $x(h)$ depending on $h$ (where $x(0)=x\in {\cal A}$).

Let the Poissonian structure, or the Poissonian bracket, on the
algebra ${\cal A}$ (the Lie bracket which is the differentiation
of ${\cal A}$), be related to the product in ${\cal B}(h)$ as
follows:
$$
\{x,y\}=\lim_{h\to 0} {i\over h} [x(h),y(h)],\qquad x,y \in {\cal A},
\quad x(h),y(h) \in {\cal B}(h)
$$
Due to polynomial dependence of the commutation relation in ${\cal
B}(h)$ on $h$ the RHS (right hand side) of the above is well
defined.

Then we say that ${\cal B}(h)$ is the quantization of ${\cal A}$,
and the linear map $x\mapsto x(h)$ is the quantization map. Note that
the quantization map is non unique in the following sense: the different
families $x(h)$ (corresponding to the different isomorphisms of linear spaces)
will reproduce the same Poissonian bracket.

This is the canonical approach to quantization, which is
sufficient to reproduce the standard models of quantum mechanics.
But this approach, in our opinion, does not take into account the
probabilistic structure of quantum mechanics and moreover, looks
too restrictive.

First, it is based on purely algebraic ideas, when quantum mechanics
is a probabilistic theory. In the discussed above standard approach we never
used the structure of the state (and any probabilistic arguments).

Second, it is obvious that in this approach one can obtain only
the commutation relations for the algebras which become
commutative for $h=0$, such as the Heisenberg algebra. It is not
possible to reproduce in the canonical approach the relations of
the quantum Boltzmann algebra (or the algebra of the free creation
and annihilation operators) of the form
$$
A_iA_j^{\dag}=\delta_{ij}
$$
which, for example, describes the statistics of collective
excitations in the quantum electrodynamics (and many other
interacting theories) \cite{book}. The relations above has no
contradiction with the celebrated spin--statistics theorem, since
they are valid for the collective interacting degrees of freedom.
This relations are derived by the time averaging procedure (called
the stochastic limit), and in \cite{book} the phenomenon of
arising of the quantum Boltzmann relations in the stochastic limit
was called the third quantization\footnote{The terminology third
quantization is quite divertive. Some authors use it in a totally
different framework.}, see Appendix for details.

It would be worth to include the mentioned above probabilistic arguments
into more general frameworks of quantization. In the present paper
we investigate the quantization procedure based on the contextual
approach to probability. We call this procedure the contextual quantization.
This quantization procedure will be relevant to a wide class of noncommutative
probability spaces with the statistics different from the Bose or Fermi
statistics, for instance for the quantum Boltzmann statistics.

In papers \cite{KHR1}--\cite{KHR4} there was developed a
contextual probabilistic approach to the statistical theory of
measurements over quantum as well as classical physical systems.
It was demonstrated that by taking into account dependence of
probabilities on complexes of experimental physical conditions,
{\it physical contexts,} we can derive quantum interference for
probabilities of alternatives. Such a contextual derivation is not
directly related to special quantum (e.g. superposition) features
of physical systems.

In the present paper we are performing the next step in developing
of the contextual probabilistic approach --- the step from the
contextual interpretation of noncommutative probability to the
contextual quantization procedure. We show that the contextual
probabilistic approach gives us not only interpretation of quantum
mechanics, but, after corresponding generalization, allows to
formulate a new quantization procedure, which is applicable, in
principle, not only to models of standard quantum mechanics, but
to a wide class of physical phenomena, which may be described by
models of noncommutative probability theory, see discussion of
\cite{nrp}, \cite{nra}, \cite{ncpro}.

To formulate the contextual quantization procedure we introduce
the following notions.

First, we introduce the general definition of probabilistic
context as a deformation of embedding of probability spaces: the
probability space ${\bf A}$ has context $f$ in the probability
space ${\bf B}$, if $f$ is the map $f: {\bf A}\to{\bf B}$ which is
a deformation of embedding of probability spaces, see Section 3
for details. This definition gives a contextual probabilistic
formulation of the correspondence principle in quantum mechanics,
but our aim is to apply it to more general situation. So the aim
of this paper is not just a new reformulation of the conventional
quantum formalism, but extension of quantum ideology to new
domains.

Second, we propose a probabilistic version of the Bohr
complementarity principle, which we call the {\it Principle of
Complementarity of Probabilities}(the PCP) and show, that this
principle is the criterion of possibility of the contextual
quantization. The Principle of Complementarity of Probabilities
describes possibility of unification of probability contexts of
several probability spaces in the frameworks of a new larger
probability space.

We discuss the relation of the introduced contextual quantization
procedure and the noncommutative replica procedure, introduced in
\cite{nrp}, \cite{nra} and show, that the noncommutative replica
procedure is the example of the contextual quantization. This
shows that the contextual quantization is applicable in the
multiplicity of the fields which is much wider than quantum
mechanics. This in some sense reflects the original ideas by Bohr.

Then we discuss applications of the contextual quantization to
collective interacting commutation relations of the stochastic
limit, see also \cite{book}, \cite{ncpro} for details.

For the review of quantum probability see \cite{Holevo}.
Quantization of thermodynamics was discussed in \cite{Maslov}.
Preliminary discussion of the mathematical framework for the
contextual approach was given in \cite{KhrVol}. Discussion of the
Chameleon point of view on quantum measurements which is similar
to contextual approach was given in \cite{Accardi},
 \cite{AR}. Contextual approach to Kolmogorov
probability with relations to quantum mechanics was discussed in
\cite{KHR2}. 

We also underline that our Principle of Complementarity of Probabilities
can be  coupled to investigations on quantum entropy
and information dynamics, see  \cite{Ohya1} -- \cite{Ohya3}.
It might be that our principle can be reformulated by using the {\it language of quantum 
entropy and information.}

The structure of the present paper is as follows.

In Section 2 we introduce the Principle of Complementarity of
Probabilities.

In Section 3 we introduce the definition of probability context as
(deformation of) embeddings of probability spaces.

In Section 4 we introduce the contextual quantization procedure
based on the notions of probability context and the Principle of
Complementarity of Probabilities.

In Section 5 we discuss the relation of the contextual
quantization and the noncommutative replica procedure of
\cite{nrp}.

In Section 6 we discuss the stochastic limit approach, in which
commutation relations for collective operators, which can not be
described by canonical quantization, but can be described by the
contextual quantization, were obtained.

\section{The Principle of Complementarity of Probabilities}

In the present Section we discuss the meaning of the
complementarity principle in quantum mechanics from the point of
view of the probabilistic interpretation of quantum mechanics and
introduce the Principle of Complementarity of Probabilities.

Probabilistic interpretation of quantum mechanics is based on the
notion of noncommutative (or quantum) probability space. A {\it
noncommutative probability space} is a pair ${\bf B}=({\cal
B},\psi)$, where ${\cal B}$, called the algebra of observables, is
the involutive algebra with unit over the complex numbers, and
$\psi$ is a state (positive normed linear functional) on ${\cal
B}$.

The commutative, or Kolmogorovian, probability space is a
particular variant of noncommutative probability space for the
case when the algebra is commutative. Such a probability space is
called classical.

The state of a quantum system is described by density matrix
--- positive functional on noncommutative algebra of observables.
Non compatible observables correspond to noncommuting operators,
which we may consider belonging to different classical
(commutative) probability subspaces in the full noncommutative
probability space. In particular, in the ordinary quantum
mechanics (where ${\cal B}$ is the Heisenberg algebra) the
position and momentum observables generate commutative subalgebras
of the Heisenberg algebra --- algebras of functions of the
position and momentum observables, respectively. Measuring the
observables from the classical subalgebra we build the (classical)
state on the classical subalgebra. Therefore, after the
observation of the full set of incompatible observables, we obtain
the set of classical states on noncommuting classical subalgebras,
or the set of classical probability spaces.

The following formulation  presented
in Discussion with Einstein on Epistemological Problems in Atomic Physics (see \cite{Bohr}, 
vol. 2, p. 40), perhaps is Bohr s most refined formulation of what he means by the
complementary situations of measurements:

{\it Evidence obtained under different experimental conditions [e.g. those of the position vs.
the momentum measurement] cannot be comprehended within a single
picture, but must be regarded as [mutually exclusive and]
complementary in the sense that only the totality of the
[observable] phenomena exhausts the possible information about the
[quantum] objects [themselves].}

We propose the following probabilistic version of the
complementarity principle. From the beginning we consider very
general model, which is essentially wider than the conventional
quantum model. Thus our aim is not only a probabilistic
reformulation of Bohr's complementarity principle, but the
extension of this principle to more general
situation\footnote{N.Bohr discussed at many occasions the
possibility to use the principle of complementarity outside of the
quantum domain \cite{Bohr}, see \cite{Plotnitsky},
\cite{Plotnitsky1}, \cite{KHR5}, \cite{KHR6}, \cite{Heisenberg1}.
However, his proposals were presented on merely philosophical
level and, as a consequence, we did not see any fruitful
applications of the principle of complementarity in other domains.
Another reason for the absence of such applications of the
principle of complementarity was Bohr's attitude to present this
principle as a kind of {\it no go} principle. For N.Bohr the main
consequence of the principle of complementarity was that {\it in
quantum mechanics, we are not dealing with an arbitrary
renunciation of a more detailed analysis of atomic phenomena, but
with a recognition that such an analysis is {\it in principle}
excluded} \cite{Bohr} (Bohr's emphasis). In the opposite, we shall
present the principle of complementarity in a constructive form by
concentrating on the possibility of {\it unification} of
statistical data obtained in incompatible experiments into a
single quantum probabilistic model. Here we are coming to the
crucial difference between Bohr's view and our view to
complementarity. Bohr's complementarity was merely {\it individual
complementarity} and our complementarity is {\it probabilistic
complementarity}. It is well known that historically N.Bohr came
to the principle of complementarity through discussion with
W.Heisenberg on his uncertainty principle. The original source of
all those considerations was the idea that for a single quantum
system the position and momentum could not be simultaneously
measured. Our main idea is that statistical data for incompatible
observables (e.g. the position and momentum) can not be obtained
in a single experiment. But, nevertheless, such data, obtained in
different experiments can be unified in a single quantum
probabilistic model. Thus, in the opposite to N.Bohr and
W.Heisenberg, we present a constructive program of the unification
of statistical data and not a {\it no go} program. Such a
constructive approach gives the possibility to extend essentially
the domain of applications of the modified principle of
complementarity.

At the same time we see the real bounds of the extension of
quantum ideology: models in which statistical data, obtained in
experiments with incompatible observables, can not be even in
principle unified into a single quantum--like probability model,
see the example at the end of this section. }.

Assume we have noncommutative involutive algebra of observables
${\cal B}$, and the set ${\cal B}_i$ of subalgebras in ${\cal B}$
with the states $\psi_i$ defined on the corresponding classical
subalgebras ${\cal B}_i$ (thus ${\bf B}_i=({\cal B}_i,\psi_i)$ are
probability spaces).

\begin{definition}\label{Def1}
{\sl We say that the states $\psi_i$ on the subalgebras ${\cal
B}_i$ of the algebra  ${\cal B}$ of observables, corresponding to
measurements of physically incompatible observables, satisfy the
Principle of Complementarity of Probabilities (or the PCP), if
they may be unified into the state $\psi$ on the full algebra of
observables ${\cal B}$. }
\end{definition}

In short the PCP may be formulated as follows: probability spaces,
corresponding to measurements of physically incompatible
observables, may be unifed into a larger probability space.

In particular, the initial probability spaces ${\bf B}_i$ can be
classical (commutative) as in the above considerations on
classical probability spaces generated by incompatible
observables. Note that in this definition we did not claim that
the initial probability spaces are necessarily commutative and
their unification is noncommutative and our definition is more
general. Therefore the PCP is not only a reformulation of the Bohr
complementarity principle in the probabilistic language, but it
also is an extension the field of applicability of the ideas of
complementarity.

\bigskip

One can see, that the PCP is not always trivially satisfied. For
example, on the algebra with the relation
$$
[a,a^*]=-1
$$
there is no positive faithful state. Therefore, classical states
on the subalgebras, generated by $a+a^*$ and $i(a-a^*)$ can not be
unified into the faithful state on the full algebra of
observables. Therefore, this algebra satisfies the original Bohr
complementarity principle but can not satisfy the Principle of
Complementarity of Probabilities.

\section{Probabilistic contexts}

In the present Section we discuss the contextual approach to
noncommutative probability and introduce the new general
definition of probabilistic context.

We remind that a morphism of probability spaces is a
$*$--homomorphism of algebras which conserves all the correlation
functions, i.e. for the morphism $f:{\bf A}=({\cal A}, \phi)\to
{\bf B}=({\cal B}, \psi)$, one has $\psi(f(a))=\phi(a)$ for any
$a\in {\cal A}$. A subspace of probability space is defined by a
subalgebra of algebra of observables and the restriction of the
state on the algebra of observables on this subalgebra. A morphism
is an embedding, if it is an injection as a $*$--homomorphism.

The contextual approach in probability theory, see
\cite{KHR1}--\cite{KHR4}, discusses the definition of probability
with respect to the complex of physically relevant conditions, or
the context. An attempt to give a formal definition of a context
was done in \cite{KhrVol}. Let us formulate the new general
definition of a context in noncommutative probability:

\begin{definition}\label{Def2}{\sl
The contextual representation (or simply context) of the
probability space ${\bf A}=({\cal A}, \phi)$ in the probability
space ${\bf B}=({\cal B}, \psi)$ is the injective
$*$--homomorphism $f:{\cal A}\to {\cal B}$, which satisfies the
correspondence principle for the states $\phi$ and $\psi$. }
\end{definition}

In the following we, if no confusion is possible, will use the
term context also for the image of $f$ in ${\bf B}$.

Two contexts are incompatible, if their images can not be
considered in the frameworks of commutative (or classical)
probability space.

Now we define (in philosofical sense) the correspondence principle,
see also \cite{KHR3} for the discussion. Identifying ${\cal A}$
with its image in ${\cal B}$, we formulate the following:

\begin{definition}\label{Def3} {\sl The states $\phi$ and $\psi$
on the involutive algebra ${\cal A}$ satisfy the correspondence
principle, if $\psi$ is a deformation of $\phi$. }
\end{definition}

Of course, the definition above does not make sense, if we do not
formulate the definition of deformation of the states. For
different algebras ${\cal A}$ and different contexts we may have
different definitions of deformation. One of the natural examples
is the following.

\begin{definition}\label{DefD} {\sl We say that the state $\psi$ on
algebra ${\cal A}$ is the deformation of the state $\phi$, if the
state $\psi=\psi_h$ depends on the (real) parameter $h$,
$\psi_0=\phi$, and for any $a\in {\cal A}$ we have $\lim_{h\to
0}\psi_h(a)=\phi(a)$.}
\end{definition}

For the prototypical example the correspondent states mean simply
equal states, and the context in Definition \ref{Def2} will be
simply an embedding of probability spaces. But this simple case
does not cover the case of standard quantum mechanics, although
may be useful for some other applications of the contextual
quantization.

\bigskip

Discuss the important examples of probability contexts, describing
the well known two slit experiment, in which we observe quantum
interference of a particle, passing through two slits to a screen.
In quantum mechanics it is naturally to define a context by fixing
of classical probability subspace (classical subalgebra with the
restriction of the state). Note that in the definition of context
we noted that we will identify the contextual map with it's image,
if no confusion is possible.

In the two slit experiment we have two important probabilistic
contexts:

\medskip

1) The context of measurements, in which we perform observations
of particles. This context, as a probability space, is generated
by projections onto the basis of measurements, and in the
considered case is given by probability space in which the
operator of coordinate along the screen is diagonal.

\bigskip

2) The context of dynamics --- probability space, in which density
matrix of the particle is diagonal.

\medskip

Since these two contexts are incompatible (i.e. density matrix of
the particle is non diagonal in the basis of measurements), we
observe quantum interference.

\section{The contextual quantization}

The introduced in Section 2 Principle of Complementarity of
Probabilities is a necessary condition for the existence of
noncommutative probability space. In the present Section we use
the Principle of Complementarity of Probabilities to define the
quantization procedure based on probabilistic arguments.

Assume we have the family of probability spaces ${\bf A}_i=({\cal
A}_i, \phi_i)$, and the family $f_{ij}:{\cal A}_i\to {\cal B}_j$
of $*$--homomorphisms of ${\cal A}_i$ onto the subalgebras ${\cal
B}_j$ of $*$--algebra ${\cal B}$, which define the states $\psi_j$
on ${\cal B}_j$:
$$
\psi_j(f_{ij}(a))=\phi_i(a)
$$
and therefore make these subalgebras the probability spaces ${\bf
B}_j=({\cal B}_j, \psi_j)$.

We propose the following definition of the contextual
quantization.

\begin{definition}\label{Def4}{\sl
Let the images ${\cal B}_j$ of algebras ${\cal A}_i$ generate the
$*$--algebra ${\cal B}$, and moreover the states $\psi_j$ on the
subalgebras ${\cal B}_j\in {\cal B}$ satisfy the Principle of
Complementarity of Probabilities and therefore there exists the
state $\psi$ on the whole $*$--algebra ${\cal B}$.

In this case we will say that the probability space ${\bf
B}=({\cal B},\psi)$ is the contextual quantization of the family
of probability spaces ${\bf A}_i$ with respect to the contexts
${\bf B}_j=f_{ij}({\bf A}_i)$.}
\end{definition}

\begin{remark}\label{pcpc}{\rm
We see that in the definition above ${\bf B}_j$ will be the
contexts of ${\bf A}_i$, and in the contextual approach the
quantization of the probability space is defined by the family of
contexts with the images satisfying the Principle of
Complementarity of Probabilities. Hence we will also call this
principle the Principle of Complementarity of Probabilistic
Contexts.}
\end{remark}

\begin{remark}\label{homomorphism}{\rm
Note that the context map $f$ in Definition \ref{Def2} is a
homomorphism, while in the canonical quantization we usually can
not use homomorphisms to map classical objects to quantum. This is
related to the fact that the context of classical probability
space in noncommutative probability space is a classical
probability subspace (and classical subalgebra (say of
coordinates) in quantum algebra may be related to classical
algebra by a homomorphism). The contextual quantization is based
on totally different idea, compared to canonical quantization:
instead of looking for classical constructions, which will be
classical traces of quantum phenomena (as in the canonical
quantization), we unify several (not necessarily classical)
statistics into more general noncommutative statistics. }
\end{remark}

Since in a general situation the morphisms of probability spaces
(and moreover their deformations) are highly non unique, the
contextual quantization is applicable in the situations which are
beyond of the frameworks of the canonical quantization. Moreover,
the contextual quantization in principle may be applied to some
special classical systems.

\bigskip

Consider the following examples. The first example describes
contextual quantization of harmonic oscillator.

\bigskip

\noindent{\bf Example 1.} \qquad Consider the classical
probability space ${\bf A}=({\cal A},\phi)$ described by the real
valued classical random variable $x$ with the Gaussian mean zero
state $\phi$, which is the Gibbs state of a classical harmonic
oscillator.

Consider the noncommutative probability space ${\bf B}=({\cal B},\psi)$,
where ${\cal B}$ is the Heisenberg algebra with the (selfadjoint) generators
$q_i$, $p_i$, $i=1,\dots,d$ and the relations
$$
[p_i,q_j]=-ih\delta_{ij}
$$
and the state $\psi$ which is the Gibbs state for the harmonic oscillator
with $d$ degrees of freedom:
$$
\psi(X)=\hbox{ tr }e^{-\beta H}X,\qquad
H={1\over 2}\sum_{i=1}^d \left(p_i^2 + q_i^2\right)
$$

Then the set of $2d$ injections \be\label{Ex1} f_i:x\mapsto
q_i,\qquad g_j:x\mapsto p_j,\qquad i,j=1,\dots,d \ee of
probability spaces ${\bf A}\to {\bf B}$ satisfies the conditions
of the Definition \ref{Def4} and defines the contextual
quantization, which transforms the classical mean zero Gaussian
real valued random variable into the quantum probability space
describing the quantum harmonic oscillator.

The correspondence principle in this example relates the Gibbs
states of classical and quantum harmonic oscillators.

Note that in the described contextual quantization, unlike in the
canonical quantization, different degrees of freedom are described
by different maps $f_i$, $g_j$ (the initial classical probability
space ${\bf A}$ has one degree of freedom). In principle we may
distinguish different degrees of freedom from the beginning and
start contextual quantization from a family of classical
probability spaces.

\bigskip

\noindent{\bf Example 2.}\qquad Consider the contextual
quantization which generates noncommutative probability space
${\bf B}=({\cal B},\psi)$, where ${\cal B}$ is the quantum
Boltzmann algebra, generated by the quantum Boltzmann
annihilations $A_i$ and creations $A^{\dag}_i$, $i=1,\dots,d$ with
the relations \be\label{qB} A_iA_j^{\dag}=\delta_{ij} \ee in the
Fock state $\psi(X)=\langle\Omega,X\Omega\rangle$, where the
vacuum $\Omega$ is annihilated by all annihilations $A_i$.

Consider the probability space ${\bf A}=({\cal A},\phi)$,
where ${\cal A}$ is the quantum Boltzmann algebra with one degree of freedom,
i.e. the algebra generated by operators $A$, $A^{\dag}$ with the relation
$$
AA^{\dag}=1
$$
and $\phi$ is the Fock state $\phi(X)=\langle\Omega,X\Omega\rangle$.
Note that the algebra ${\cal A}$ is noncommutative.

Consider the set of $d$ embeddings of probability spaces ${\bf
A}\to {\bf B}$ \be\label{Ex2} A\mapsto A_i,\qquad A^{\dag}\mapsto
A_i^{\dag},\qquad i=1,\dots,d \ee This set of injections satisfies
the conditions of Definition \ref{Def4} (where the deformation in
the correspondence principle is an identity, i.e. the
correspondent states are identically equal) and therefore the
probability space ${\bf B}$ (quantum Boltzmann for $d$ degrees of
freedom in the Fock state) is the contextual quantization of the
probability space ${\bf A}$ (the same for one degree of freedom).

\bigskip

\noindent{\bf Example 3.}\qquad In the Example 2 we quantized the
noncommutative probability space ${\bf A}$ and obtained the more
complicated noncommutative probability space ${\bf B}$. Note that
if we consider in the noncommutative probability space ${\bf A}$
the commutative subspace ${\bf A}_0$ with the algebra of
observables generated by $X=A+A^{\dag}$ and consider the
restriction of the quantization procedure on ${\bf A}_0$: we
consider the embeddings ${\bf A}_0\to {\bf B}_0$, \be\label{Ex20}
X\mapsto X_i=A_i+A_i^{\dag},\qquad i=1,\dots,d \ee where ${\bf
B}_0$ is the probability space with the algebra of observables
generated by $X_i$ in the Fock state, then the contextual
quantization ${\bf B}_0$ of noncommuting contexts of commutative
probability space ${\bf A}_0$ will be a noncommutative probability
space.

\bigskip

For further discussion of relations used in Examples 2 and 3 see
Appendix.

\bigskip

\noindent{\bf Example 4.}\qquad Example 1 can be generalized to
quantization in superspace, see \cite{KHRsup}. We would not like
to go into details (since it needs a few new definitions which are
not directly relevant to contextual quantization). We just mention
that there is considered a quantization of supercommutative
superalgebras. Thus initial classical algebras are not
commutative, but supercommutative.

\bigskip

The next example describes that one can call a {\it cognitive
quantization}.

\bigskip

\noindent{\bf Example 5.}\qquad The next example has no direct
relation to quantum mechanics, but we hope, it will find
applications in the future. Assume we have several persons which
have different points of view on some complex problem. Each point
of view is described by a classical probability space, which
describes the distribution of possible opinions. If the problem
under consideration is complex enough, it may be possible that
different points of view can not be unified within the frameworks
of a single point of view, or in our description, within the
frameworks of a single commutative probability space. Instead, it
may be possible, that different points of view (or corresponding
probability spaces) are complementary and may be unified within
the frameworks of noncommutative probability space, which gives a
relevant complete description of the considered complex problem,
while it is not possible to give such a description using
classical probability space (or single point of view). The
Principle of Complementarity of Probabilities here guarantees,
that there is no contradiction between different points of view
and these points of view are complementary but not contradictory.

\bigskip

In the next Section we discuss the contextual quantization in
relation to the noncommutative replica approach, introduced in
\cite{nrp}, \cite{nra}, which was one of the motivations for the
definitions of the present paper.

\section{The contextual quantization and replicas}

In the present Section we show that the noncommutative replica procedure
for disordered systems is the example of the contextual quantization
and the contexts describe the way of averaging of quenched disorder.

In papers \cite{nrp}, \cite{nra} the noncommutative replica
procedure for disordered systems was introduced. The
noncommutative replica procedure has the form of the embedding of
probability spaces. It was mentioned that this embedding is non
unique and moreover the combination of different morphisms can not
be considered in a commutative probability space, see \cite{nrp},
\cite{ncpro}.

The noncommutative replica procedure is defined (by S. Kozyrev \cite{nrp}, \cite{nra}) as follows.
Consider the system of random Gaussian $N\times N$ matrices
$J_{ij}$ with independent matrix elements with the zero mean and
unit dispersion of each of the matrix element. Consider disordered
system with the Hamitlonian $H[\sigma,J]$, where $\sigma$
enumerates the states of the system and the disorder $J$ is the
mentioned above random matrix.

Introduce  the commutative probability space where random variable
corresponds to the random matrix $J$ and the correlations are defined as
by the correlator
\be\label{corr}
\langle J^k \rangle= Z^{-1}
{1\over N^k} E\left(\sum_{\sigma}e^{-\beta H[\sigma,J]}\hbox{ tr } J^k\right)
\ee
where $E$ is the expectation of the matrix elements and
$$
Z=E\left(\sum_{\sigma}e^{-\beta H[\sigma,J]}\right)
$$
This probability space describes annealed disordered system
and the averaging over $J$ describes the averaging of the disorder.
For $\beta=0$ (\ref{corr})
reduces to the Gaussian state $E$ on random matrices.

We consider the family of contexts of the annealed probability
space in the larger replica probability space generated by the
replicas $J_{ij}^{(a)}$ of the random matrix $J_{ij}$, see
\cite{nrp} for details. These contexts we call the noncommutative
replica procedures and they have the form \be\label{Delta} \Delta:
J_{ij}\mapsto {1\over\sqrt{p}}\sum_{a=0}^{p-1} c_a J_{ij}^{(a)};
\ee This embedding of probability spaces describes the way of
selfaveraging of the quenched disorder.

The context $\Delta$ maps the matrix element $J_{ij}$ into the linear
combination of independent replicas $J_{ij}^{(a)}$, enumerated by
the replica index $a$.
Here $c_a$ are complex coefficients, which should satisfy the
condition
$$
\sum_{a=0}^{p-1} |c_a|^2 =p
$$
Varying coefficients $c_a$ we will obtain different morphisms
$\Delta$ of probability spaces.

After the morphism $\Delta$ the probability space is described by
the correlation functions (\ref{corr}) with $J$ replaced by
$\Delta J$ and mathematical expectation $E$ replaced by the
analogous expectation for the set of independent random matrices
(replicas of $J$).

Discuss the large $N$ limit of the probability space (\ref{corr})
for the free case $\beta=0$.

In the large $N$ limit, by the Wigner theorem, see
\cite{Wig}--\cite{ALV}, the system of $p$ random matrices with
independent variables will give rise to the quantum Boltzmann
algebra with $p$ degrees of freedom with the generators $A_a$,
$A_a^{\dag}$, $a=0,\dots,p-1$ and the relations
$$
A_aA_b^{\dag}=\delta_{ab}
$$
The Gaussian state on large random matrices in the $N\to\infty$ limit
becomes the Fock, or vacuum, state on the quantum Boltzmann algebra:
the Fock state is generated by the expectation
$\langle\Omega, X\Omega\rangle$
where $\Omega$ is the vacuum: $A_a\Omega=0$, $\forall a$.

The operators $A_a$ are the limits of the large random matrices
$$
\lim_{N\to\infty}{1\over N}J_{ij}^{(a)}=Q_a=A_a+A_a^{\dag}
$$
where the convergence is understood in the sense of correlators
(as in the central limit theorem).

Then in the thermodynamic limit $N\to\infty$ the map
(\ref{Delta}) will take the form of the following map
(for which we use the same notation) of the quantum Boltzmann algebra
with one degree of freedom into quantum Boltzmann algebra with $p$ degrees of
freedom:
$$
\Delta: A\mapsto {1\over\sqrt{p}}\sum_{a=0}c_a A_a
$$
and correspondingly
$$
\Delta: Q\mapsto {1\over\sqrt{p}}\sum_{a=0}c_a Q_a
$$

We see that the set of contexts $\Delta$ with the different
coefficients $c_a$ defines the contextual quantization of the commutative
probability space, described in (\ref{Ex20}).

\section{Appendix: the Stochastic Limit}

In the present Section we briefly discuss the stochastic limit
approach, in which deformations of quantum Boltzmann commutation
relations were obtained, see \cite{book}. In this approach we
consider the quantum system with the Hamiltonian in the form
$$
H=H_0+\lambda H_I
$$
where $H_0$ is called the free Hamiltonian, $H_I$ is called the
interaction Hamiltonian, and $\lambda\in {\bf R}$ is the coupling
constant.

We investigate the dynamics of the system in the new slow time
scale of the stochastic limit, taking the van Hove time rescaling
\cite{vanHove}
$$
t\mapsto t/\lambda^2
$$
and considering the limit $\lambda\to 0$. In this limit the free
evolutions of the suitable collective operators
$$
A(t,k)=e^{itH_0}A(k)e^{-itH_0}
$$
will become quantum white noises:
$$
\lim_{\lambda\to 0}{1\over\lambda}A\left({t\over
\lambda^2},k\right)=b(t,k)
$$
The convergence is understood in the sense of correlators. The
$\lambda\to 0$ limit describes the time averaging over
infinitesimal intervals of time and allows to investigate the
dynamics on large time scale, where the effects of interaction
with the small coupling constant $\lambda$ are important. For the
details of the procedure see \cite{book}.

The collective operators describe joint excitations of different
degrees of freedom in systems with interaction, and may have the
form of polynomials over creations and annihilation of the field,
or may look like combinations of the field and particles operators
etc.

For example, for nonrelativistic quantum electrodynamics without
the dipole approximation the collective operator is \be\label{QED}
A_j(k)=e^{ikq}a_j(k) \ee where $a_j(k)$ is the annihilation of the
electromagnetic (Bose) field with wave vector $k$ and polarization
$j$, $q=(q_1,q_2,q_3)$ is the position operator of quantum
particle (say electron), $qk=\sum_{i}q_ik_i$.

The nontrivial fact is that, after the $\lambda\to 0$ limit,
depending on the form of the collective operator, the statistics
of the noise $b(t,k)$ depends on the form of the collective
operator and may be nontrivial.

Consider the following examples.

1) We may have the following possibility \be\label{Bose}
[b_i(t,k),b_j^{\dag}(t',k')]=2\pi\delta_{ij}\delta(t-t')\delta(k-k')\delta(\omega(k)-\omega_0)
\ee which corresponds to the quantum electrodynamics in the dipole
approximation, describing the interaction of the electromagnetic
field with two level atom with the level spacing (energy
difference of the levels) equal to $\omega_0$. Here $\omega(k)$ is
the dispersion of quantum field.

In this case the quantum noise will have the Bose statistics, and
different annihilations of the noise will commute
$$
[b_i(t,k),b_j(t',k')]=0
$$

2) The another possibility is the relation \be\label{qB1}
b_i(t,k)b_j^{\dag}(t',k')=2\pi\delta_{ij}\delta(t-t')\delta(k-k')\delta(\omega(k)+\varepsilon(p)-\varepsilon(p+k))
\ee which corresponds to the quantum electrodynamics without the
dipole approximation with interacting operator (\ref{QED}). Here
$\omega(k)$ and $\varepsilon(p)$ are dispersion functions of the
field and of the particle correspondingly.

In this case the quantum noise will have the quantum Boltzmann
statistics \cite{book},  \cite{QED}, and different
annihilations of the noise will not commute
$$
b_i(t,k) b_j(t',k')\ne b_j(t',k') b_i(t,k)
$$

The commutation relations of the types (\ref{Bose}), (\ref{qB1})
are universal in the stochastic limit approach (a lot of systems
will have similar relations in the stochastic limit $\lambda\to 0
$).

\section{Appendix: Copenhagen and V\"axj\"o  complementarities}

In the first vesrion of this paper, see \cite{KOZKHR}, we used the following version of the 
complementarity principle (V\"axj\"o  complementarity): 

{\it To obtain the full information about the state of a quantum system,
one has to perform measurements of the set of physically 
incompatible (noncommuting) observables, say the momentum and the
coordinate. Measuring only the momentum or only the coordinate we
will not obtain the full information about the state of the
quantum system.}

In his Email to one of the authors (A. Yu. Khrennikov) A. Plotnitsky remarked:

``First of all, this formulation does not appear to me to correspond to Bohr's 
view to complementarity and indeed implies a conflict bewteen the respective interpretations
of quantum mechanics, Bohr's and your own\footnote{A. Plotnitsky mentioned the so called V\"axj\"o 
interpretation of quantum mechanics, see \cite{VI}, \cite{VI1}, cf. \cite{CH}, 
\cite{CH1}.}, or any interpretation consistent with your formulation...''

We agree with A. Plotnitsky and in this paper we use the original Bohr's formulation. The difference 
between two formulations, ``Copenhagen and V\"axj\"o  complementarities'', is that N. Bohr considered
{\it possible information} and we considered {\it full information.} To consider  ``possible information''
one need not to use a realists interpretation  and to consider ``full information'' (about something) we need 
to use a realists interpretation, e.g., the V\"axj\"o 
interpretation.

However, in this paper we do not try to connect our mathematical formulation of an 
extended principle of complementarity to any fixed interpretation of quantum mechanics.
This is just a formal mathematically formalized principle. This principle can be considered
as the formalization of either the Bohr's principle or V\"axj\"o principle. 

Finally, we remark that problems discussed in this paper are closely related to the problem
of understanding of information in quantum theory, see \cite{CH}-- \cite{CH3}, \cite{Plotnitsky}.

\bigskip

\centerline{\bf Acknowledgements}

The authors would like to thank L.Accardi and I.V.Volovich, L.
Ballentine, S. Albeverio, S. Gudder, W. De Muynck, J. Summhammer,
P. Lahti, A. Holevo,  B. Hiley for fruitful (and rather critical)
discussions.

\end{document}